\documentclass[12pt]{article}

\usepackage[letterpaper, margin=1in]{geometry}

\usepackage[colorlinks]{hyperref}
\usepackage{makeidx}
\usepackage{framed}
\usepackage{fancyvrb}
\usepackage{color}
\usepackage{hyperref}
\hypersetup{colorlinks = true, allcolors = blue}
\usepackage{float}
\usepackage{flafter}
\usepackage{tabularx}
\usepackage{booktabs}
\usepackage{longtable}
\usepackage{multirow}
\usepackage{graphicx}
\newcommand{\pandocbounded}[1]{#1}
\usepackage{afterpage}

\usepackage{datetime}

\yyyymmdddate

  \usepackage{setspace}
  \AtBeginEnvironment{tabular}{\singlespacing}
  \AtBeginEnvironment{table}{\singlespacing}
  \AtBeginEnvironment{longtable}{\singlespacing}
  \AtBeginEnvironment{verbatim}{\singlespacing}
  \AtBeginEnvironment{Shaded}{\singlespacing}

\NewDocumentCommand\citeproctext{}{}

\makeatletter
 \let\@cite@ofmt\@firstofone
 \def\@biblabel#1{}
 \def\@cite#1#2{{#1\if@tempswa , #2\fi}}
\makeatother
\newlength{\cslhangindent}
\newlength{\csllabelwidth}
\newenvironment{CSLReferences}[2] % #1 hanging-indent, #2 entry-spacing
 {\begin{list}{}{%
  \setlength{\itemindent}{0pt}
  \setlength{\leftmargin}{0pt}
  \setlength{\parsep}{0pt}
  \ifodd #1
   \setlength{\leftmargin}{\cslhangindent}
   \setlength{\itemindent}{-1\cslhangindent}
  \fi
  \setlength{\itemsep}{#2\baselineskip}}}
 {\end{list}}
\usepackage{calc}

  \usepackage{amsmath}
  \usepackage{amssymb}

\usepackage{xcolor}
\definecolor{dark}{HTML}{2c2e35}
\color{dark}

\definecolor{myblue}{HTML}{1e3765}
\hypersetup{colorlinks,linkcolor=myblue,urlcolor=myblue}

\makeindex

\providecommand{\tightlist}{\setlength{\itemsep}{0pt}\setlength{\parskip}{0pt}}

  \usepackage{color}
  \usepackage{fancyvrb}

  \DefineVerbatimEnvironment{Highlighting}{Verbatim}{commandchars=\\\{\}}
  \usepackage{framed}
  \definecolor{shadecolor}{RGB}{241,243,245}
  \newenvironment{Shaded}{\begin{snugshade}}{\end{snugshade}}

  \newcommand{\AttributeTok}[1]{\textcolor[rgb]{0.40,0.45,0.13}{#1}}

  \newcommand{\CommentTok}[1]{\textcolor[rgb]{0.37,0.37,0.37}{#1}}

  \newcommand{\DecValTok}[1]{\textcolor[rgb]{0.68,0.00,0.00}{#1}}

  \newcommand{\FunctionTok}[1]{\textcolor[rgb]{0.28,0.35,0.67}{#1}}

  \newcommand{\NormalTok}[1]{\textcolor[rgb]{0.00,0.23,0.31}{#1}}
  
  \newcommand{\OtherTok}[1]{\textcolor[rgb]{0.00,0.23,0.31}{#1}}

  \newcommand{\SpecialCharTok}[1]{\textcolor[rgb]{0.37,0.37,0.37}{#1}}
  
  \newcommand{\StringTok}[1]{\textcolor[rgb]{0.13,0.47,0.30}{#1}}

\author{
  Mauricio Vargas Sepúlveda\\Department of Political Science, University
of Toronto\\Munk School of Global Affairs and Public Policy, University
of Toronto\\
  \smallskip\\
  \smallskip\\
  Corresponding author: m.sepulveda@mail.utoronto.ca
}
\title{kendallknight: An R Package for Efficient Implementation of
Kendall's Correlation Coefficient Computation}
\date{Last updated: \today\ \currenttime}

\begin{document}

\maketitle

\thispagestyle{empty}
\tableofcontents
\setcounter{page}{0}
\clearpage

\afterpage{\setlength\parskip{10pt}}

\section{Abstract}\label{abstract}

The kendallknight package introduces an efficient implementation of
Kendall's correlation coefficient computation, significantly improving
the processing time for large datasets without sacrificing accuracy. The
kendallknight package, following Knight (1966) and posterior literature,
reduces the computational complexity resulting in drastic reductions in
computation time, transforming operations that would take minutes or
hours into milliseconds or minutes, while maintaining precision and
correctly handling edge cases and errors. The package is particularly
advantageous in econometric and statistical contexts where rapid and
accurate calculation of Kendall's correlation coefficient is desirable.
Benchmarks demonstrate substantial performance gains over the base R
implementation, especially for large datasets.

\section{Introduction}\label{introduction}

Kendall's correlation coefficient is a non-parametric measure of
association between two variables and it is particularly useful to
compute pseudo-\(R^2\) statistics in the context of Poisson regression
with fixed effects (Silva and Tenreyro 2006).

The current Kendall's correlation coefficient implementation in R has a
computational complexity of \(O(n^2)\), which can be slow for large
datasets (R Core Team 2024). While R features a highly efficient
multi-threaded implementation of the Pearson's correlation coefficient,
the Kendall's case that is also multi-threaded can be particularly slow
for large datasets (e.g.~10,000 observations or more).

We used C++ in the \texttt{kendallknight} package to compute the
Kendall's correlation coefficient in a more efficient way, with a
computational complexity of \(O(n \log(n))\), following Knight (1966),
Abrevaya (1999), Christensen (2005) and Emara (2024).

For a dataset with 20,000 observations, a computational complexity
\(O(n^2)\) involves 400 million operations and a computational
complexity \(O(n \log(n))\) requires approximately 198,000 operations to
obtain the Kendall's correlation coefficient.

Our implementation can reduce computation time by several minutes or
hours as we show in the benchmarks, and without sacrificing precision or
correct handling of corner cases as we verified with exhaustive testing.

\section{Definitions}\label{definitions}

Kendall's correlation coefficient is a pairwise measure of association
and it does not require the data to be normally distributed. For two
vectors \(x\) and \(y\) of length \(n\), it is defined as (Knight 1966):

\begin{equation*}
r(x,y) = \frac{c - d}{\sqrt{(c + d + e)(c + d + f)}},
\end{equation*}

where \(c\) is the number of concordant pairs, \(d\) is the number of
discordant pairs, \(e\) is the number of ties in \(x\) and \(f\) is the
number of ties in \(y\).

The corresponding definitions for \(c\), \(d\), \(e\) and \(f\) are:

\begin{eqnarray*}
c &=& \sum_{i=1}^{n} \sum_{j \neq i}^{n} g_1(x_i, x_j, y_i, y_j), \\
d &=& \sum_{i=1}^{n} \sum_{j \neq i}^{n} g_2(x_i, x_j, y_i, y_j), \\ 
e &=& \sum_{i=1}^{n} \sum_{j \neq i}^{n} g_3(x_i, x_j) g_4(y_i, y_j), \\
f &=& \sum_{i=1}^{n} \sum_{j \neq i}^{n} g_4(x_i, x_j) g_3(y_j, y_i).
\end{eqnarray*}

The functions \(g_1\), \(g_2\), \(g_3\) and \(g_4\) are indicators
defined as:

\begin{eqnarray*}
g_1(x_i, x_j, y_i, y_j) &=& \begin{cases}
  1 & \text{if } (x_i - x_j)(y_i - y_j) > 0, \\
  0 & \text{otherwise},
\end{cases} \\
g_2(x_i, x_j, y_i, y_j) &=& \begin{cases}
  1 & \text{if } (x_i - x_j)(y_i - y_j) < 0, \\
  0 & \text{otherwise},
\end{cases} \\
g_3(x_i, x_j) &=& \begin{cases}
  1 & \text{if } x_i = x_j \text{ and } y_i \neq y_j, \\
  0 & \text{otherwise},
\end{cases} \\
g_4(y_i, y_j) &=& \begin{cases}
  1 & \text{if } x_i \neq x_j \text{ and } y_i = y_j, \\
  0 & \text{otherwise}.
\end{cases}
\end{eqnarray*}

Kendall's correlation correlation is a measure of the proportion of
concordant pairs minus the proportion of discordant pairs corrected by
the proportion of ties in the data, and it requires to compare
\(m = n(n - 1) / 2\) pairs of observations which is why its
computational complexity is \(O(n^2)\).

Without ties, or duplicates in the data, the Kendall's correlation
coefficient simplifies to:

\begin{equation*}
r(x,y) = \frac{c - d}{c + d} = 
 \frac{c - d}{m} =
 % \frac{2(c - d)}{n(n - 1)} = 
 % \frac{2c}{n(n - 1)} - \frac{2d}{n(n - 1)} =
 % \frac{2c}{n(n - 1)} - \frac{2(m - c)}{n(n - 1)} =
 % \frac{4c}{n(n - 1)} - \frac{2m}{n(n - 1)} =
 % \frac{4c}{n(n - 1)} - \frac{2m}{2m} =
 \frac{4c}{n(n - 1)} - 1
\end{equation*}

A naive implementation consisting in comparing all pairs of observations
has a computational complexity of \(O(n^2)\). However, the Kendall's
correlation coefficient can be computed more efficiently by sorting the
data and using the number of inversions in the data to compute the
correlation with a computational complexity of \(O(n \log(n))\) using
binary trees (Knight 1966).

\section{Implementation}\label{implementation}

Using a merge sort with a binary tree with a depth \(1 + \log_2(n)\)
results in a search and insert operation with a computational complexity
of \(O(\log(n))\), resulting in a computational complexity of
\(O(n \log(n))\) for the Kendall's correlation coefficient (Knight 1966;
Emara 2024).

An algorithm that conducts the following operations can compute the
Kendall's correlation coefficient in an efficient way, with
computational complexity of \(O(n \log(n))\) instead of \(O(n^2)\), as
follows:

\begin{enumerate}
\def\labelenumi{\arabic{enumi}.}
\tightlist
\item
  Sort the vector \(x\) and keep track of the original indices in a
  permutation vector.
\item
  Rearrange the vector \(y\) according to \(x\).
\item
  Compute the total pairs \(m\).
\item
  Compute the pairs of ties in \(x\) as \(m_x = t_x (t_x + 1) / 2\).
\item
  Compute the pairs of ties in \(y\) as \(m_y = t_y (t_y + 1) / 2\).
\item
  Compute the concordant pairs adjusted by the number of swaps in \(y\)
  by using a merge sort as \(t = m - t_x - t_y + 2t_p\).
\item
  Compute the Kendall's correlation coefficient as
  \(r(x,y) = t / (\sqrt{m - m_x} \sqrt{m - m_y})\).
\end{enumerate}

The \texttt{kendallknight} package implements these steps in C++ and
exports the Kendall's correlation coeeficient as a function that can be
used in R by using the \texttt{cpp11} headers (Vaughan, Hester, and
François 2023). Unlike existing implementations with \(O(n \log(n))\)
complexity, such as Filzmoser, Fritz, and Kalcher (2023), this
implementation also provides dedicated functions to test the statistical
significance of the computed correlation, and for which it uses a C++
port of the Gamma function that R already implemented in C (R Core Team
2024).

\section{Benchmarks}\label{benchmarks}

We tested the \texttt{kendallknight} package against the base R
implementation of the Kendall correlation using the \texttt{cor}
function with \texttt{method\ =\ "kendall"} for randomly generated
vectors of different lengths. The results are shown in the following
tables:

\begin{longtable}[]{@{}
  >{\raggedright\arraybackslash}p{(\linewidth - 4\tabcolsep) * \real{0.3026}}
  >{\raggedleft\arraybackslash}p{(\linewidth - 4\tabcolsep) * \real{0.3947}}
  >{\raggedleft\arraybackslash}p{(\linewidth - 4\tabcolsep) * \real{0.3026}}@{}}
\toprule\noalign{}
\begin{minipage}[b]{\linewidth}\raggedright
Number of observations
\end{minipage} & \begin{minipage}[b]{\linewidth}\raggedleft
kendallknight median time (s)
\end{minipage} & \begin{minipage}[b]{\linewidth}\raggedleft
base R median time (s)
\end{minipage} \\
\midrule\noalign{}
\endhead
\bottomrule\noalign{}
\endlastfoot
10,000 & 0.003 & 1.251 \\
20,000 & 0.010 & 5.313 \\
30,000 & 0.011 & 11.002 \\
40,000 & 0.014 & 19.578 \\
50,000 & 0.017 & 30.509 \\
60,000 & 0.021 & 43.670 \\
70,000 & 0.024 & 61.310 \\
80,000 & 0.029 & 77.993 \\
90,000 & 0.031 & 98.614 \\
100,000 & 0.035 & 121.552 \\
\end{longtable}

\begin{longtable}[]{@{}
  >{\raggedright\arraybackslash}p{(\linewidth - 4\tabcolsep) * \real{0.2556}}
  >{\raggedleft\arraybackslash}p{(\linewidth - 4\tabcolsep) * \real{0.4111}}
  >{\raggedleft\arraybackslash}p{(\linewidth - 4\tabcolsep) * \real{0.3333}}@{}}
\toprule\noalign{}
\begin{minipage}[b]{\linewidth}\raggedright
Number of observations
\end{minipage} & \begin{minipage}[b]{\linewidth}\raggedleft
kendallknight memory allocation (MB)
\end{minipage} & \begin{minipage}[b]{\linewidth}\raggedleft
base R memory allocation (MB)
\end{minipage} \\
\midrule\noalign{}
\endhead
\bottomrule\noalign{}
\endlastfoot
10,000 & 1.257 & 0.812 \\
20,000 & 2.061 & 1.450 \\
30,000 & 3.091 & 2.175 \\
40,000 & 4.121 & 2.900 \\
50,000 & 5.151 & 3.625 \\
60,000 & 6.181 & 4.350 \\
70,000 & 7.211 & 5.074 \\
80,000 & 8.241 & 5.799 \\
90,000 & 9.271 & 6.524 \\
100,000 & 10.301 & 7.249 \\
\end{longtable}

These results can be complements with the following plots:

\pandocbounded{\includegraphics[keepaspectratio]{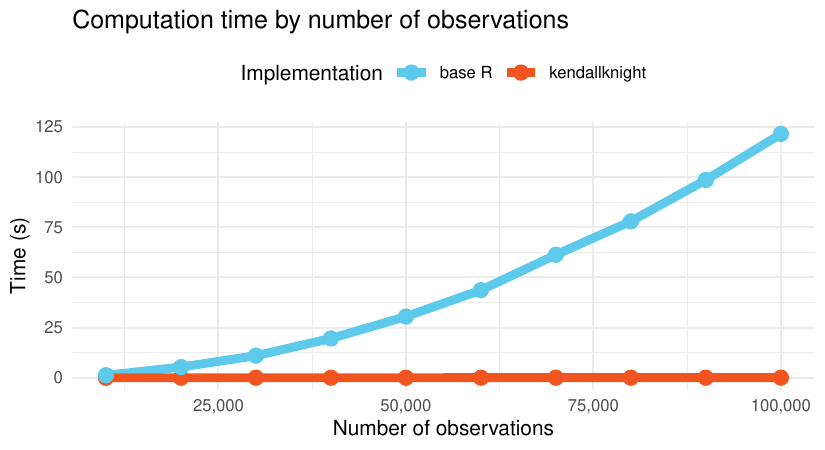}}

\pandocbounded{\includegraphics[keepaspectratio]{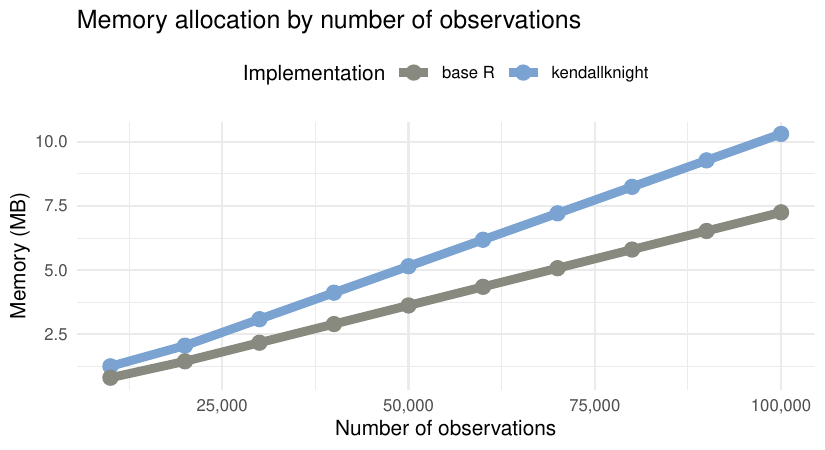}}

As a reference, estimating the coefficients for a Poisson regression
using a dataframe with around 28,000 observations, five variables and
around 700 exporter-time and importer-time fixed effects takes around
0.4 seconds with the \texttt{capybara} package (Vargas Sepulveda 2024).

To obtain summary tables for the same model, including clustered
standard errors, significance and pseudo-\(R^2\), it takes around 0.2
additional seconds using \texttt{kendallknight}. Using base R to compute
the Kendall's correlation coefficient for the pseudo-\(R^2\) statistic
takes around 7.5 seconds without the rest of statistics.

The benchmarks were conducted on a ThinkPad X1 Carbon Gen 9 with the
following specifications:

\begin{itemize}
\tightlist
\item
  Processor: Intel Core i7-1185G7 with eight cores
\item
  Memory: 16 GB LPDDR4Xx-4266
\item
  Operating System: Pop!\_OS 22.04 based on Ubuntu 22.04
\item
  R Version: 4.4.1
\item
  BLAS Library: OpenBLAS 0.3.20
\end{itemize}

\section{Testing}\label{testing}

The package uses \texttt{testthat} for testing (Wickham 2011). The
included tests are exhaustive and covered the complete code to check for
correctness comparing with the base R implementation, and also checking
corner cases and forcing errors by passing unusable input data to the
user-visible functions. The current tests cover 100\% of the code.

\section{Installation and Usage}\label{installation-and-usage}

The \texttt{kendallknight} package is available on CRAN and can be
installed using the following command:

\begin{Shaded}
\begin{Highlighting}[]
\CommentTok{\# CRAN}
\FunctionTok{install.packages}\NormalTok{(}\StringTok{"kendallknight"}\NormalTok{)}

\CommentTok{\# GitHub}
\NormalTok{remotes}\SpecialCharTok{::}\FunctionTok{install\_github}\NormalTok{(}\StringTok{"pachadotdev/kendallknight"}\NormalTok{)}
\end{Highlighting}
\end{Shaded}

The package can be used as in the following example:

\begin{Shaded}
\begin{Highlighting}[]
\FunctionTok{library}\NormalTok{(kendallknight)}

\FunctionTok{set.seed}\NormalTok{(}\DecValTok{200}\NormalTok{)}
\NormalTok{x }\OtherTok{\textless{}{-}} \FunctionTok{rnorm}\NormalTok{(}\DecValTok{100}\NormalTok{)}
\NormalTok{y }\OtherTok{\textless{}{-}} \FunctionTok{rnorm}\NormalTok{(}\DecValTok{100}\NormalTok{)}

\FunctionTok{kendall\_cor}\NormalTok{(x, y)}
\end{Highlighting}
\end{Shaded}

\begin{verbatim}
[1] 0.1288889
\end{verbatim}

\begin{Shaded}
\begin{Highlighting}[]
\FunctionTok{kendall\_cor\_test}\NormalTok{(x, y, }\AttributeTok{alternative =} \StringTok{"less"}\NormalTok{)}
\end{Highlighting}
\end{Shaded}

\begin{verbatim}
$statistic
[1] 0.1288889

$p_value
[1] 0.971286

$alternative
[1] "alternative hypothesis: true tau is less than 0"
\end{verbatim}

\section{Conclusion}\label{conclusion}

The \texttt{kendallknight} package provides a fast and memory-efficient
implementation of the Kendall's correlation coefficient with a
computational complexity of \(O(n \log(n))\), which is orders of
magnitude faster than the base R implementation without sacrificing
precision or correct handling of corner cases. For small vectors (e.g.,
less than 100 observations), the time difference is negligible. However,
for larger vectors, the difference can be substantial. This package is
particularly useful to solve bottlenecks in the context of econometrics
and international trade, but it can also be used in other fields where
the Kendall's correlation coefficient is required.

\section*{References}\label{references}
\addcontentsline{toc}{section}{References}

\phantomsection\label{refs}

% Index?
% \printindex

\end{document}